\documentclass[aps,reprint]{revtex4-1}
\usepackage{blindtext}

\usepackage{amsfonts}
\usepackage{color}
\usepackage{graphicx}
\usepackage{braket}
\usepackage{dcolumn}
\usepackage{bm}
\usepackage{subfigure}
\usepackage{amssymb}
\usepackage{multirow}
\usepackage{natbib}
\usepackage{amsmath}
\usepackage{mathtools}
\usepackage{xcolor}

\usepackage{sansmath}
\usepackage{txfonts}

\begin{document}

\title{Modification of 
the uniform electron gas polarizational stopping power \\ due to the interaction of the projectile with new collective modes at moderate and strong coupling}

\author{
S.A. Syzganbayeva$^{1}$, A.V. Filinov$^{1,2\star}$, \fbox{Jes\'{u}s Ara}$^{3}$, A.B. Ashikbayeva$^{1}$, 
A. Askaruly$^{1}$, L.T. Yerimbetova$^{1}$, \\ M.D. Barriga-Carrasco$^{4}$, Y.V. Arkhipov$^{1}$, I.M. Tkachenko$^{5,1\star}$
}

\address{$^{1}$Al-Farabi Kazakh National University, Almaty, Kazakhstan}
\address{$^{2}$Institut für Theoretische Physik und Astrophysik, Christian-Albrechts-Universität zu Kiel, Germany}
\address{$^{3}$Instituto de Tecnolog\'{\i}a Qu\'{\i}mica, Universitat Polit\`{e}cnica de Val\`{e}ncia,Spain}
\address{$^{4}$ ETSI Industrial de Ciudad Real Universidad de Castilla-La Mancha, Ciudad Real, Spain}

\address{$^{5}$ Departament de Matem\`{a}tica Aplicada, Universitat Poli\`{e}cnica de Val\`{e}ncia, Valencia, Spain}

\email{imtk@upv.edu.es, al.filinov@gmail.com}

\date{02.05.2025}

\begin{abstract}
This paper presents a detailed study of the polarizational stopping power of a homogeneous electron gas in moderate and strong coupling regimes using the self-consistent version of the method of moments as the key theoretical approach capable of expressing the dynamic characteristics of the system in terms of the static ones, which are the moments. We develop a robust framework that relies on nine sum rules and other exact relationships
to analyze electron-electron interactions and their impact on energy-loss processes. 
We derive an expression for the stopping power that takes into account both quantum
statistical effects and electron correlation phenomena. Our results demonstrate significant deviations from classical stopping power predictions, especially under the strong coupling conditions when electron dynamics is highly dependent on collective behavior and a projectile interacts with the system collective modes revealed in Phys. Rev. B $\bf{107}$, 195143 (2023). This work not only advances the theoretical understanding of the homogeneous electron gas but also has implications for practical applications in fields such as plasma physics and materials science.

\end{abstract}

\maketitle

\section{Introduction}\label{sec1}

The uniform electron gas (UEG) serves as a fundamental model in many-body
physics, providing important insights into the behavior of electrons in
systems ranging from plasmas to metals. Understanding the UEG polarizational
stopping power, defined as the energy loss per unit length experienced by
charged particles passing through a medium that neglects the projectile's
feedback to the latter, is vital for applications in nuclear and plasma physics,
astrophysics, and materials science. Much of the work has been done, especially using the $\it{ab initio}$ Quantum Monte-Carlo method,  
on the UEG dynamic properties in the weak-coupling regime, but a significant gap remains in our
understanding of this system behavior in moderate to strong coupling conditions, where
electronic interaction becomes more pronounced and collective effects play a
critical role. Recent achievements are reflected in \cite{dornheim1, dornheim2, Bellenbaum} and references therein, to name a few.

Alternatively, we use the non-perturbative self-consistent method of
moments, taking into account nine sum rules and other exact relations, a
powerful theoretical tool that permits to investigate the UEG dynamic properties in
general. The moment approach allows us to systematically explore the
statistical and dynamic properties of the electron gas by examining the
moments of a tailored distribution function, which makes it easier to analyze
complex interactions in a more understandable way. By focusing on the interaction strength, we aim to derive expressions that capture the rich interplay between
electron density, temperature, and quantum and correlation effects, going beyond the random-phase, Mermin, or FCDF (Full Conserving Dielectric Function) approximations proved, at least under some specific conditions, to lead to very similar conclusions with respect to the UEG stopping power and straggling; see \cite{Barriga-Carrasco_2011,PhysRevE.82.046403} and references therein. Our study is motivated by the need to link theoretical predictions with simulation results
and even experimental observations, especially in regimes where standard models are unable to capture the intricacies of electron dynamics. We provide a comprehensive framework that not only improves theoretical understanding,
but also has practical implications for the design and analysis of materials
in various technological and scientific fields. This paper presents our
findings, contributing to a deeper understanding of the behavior of electrons in
many-body systems and predicting the plasma stopping and friction function under extreme conditions otherwise difficult to attain by alternative methods and approaches used nowadays in the field of plasma physics and condensed matter.

\section{The problem set-up
}
\label{sec2}
The present approach is used to calculate the dynamic
characteristics of non-perturbative systems, taking into account nine sum
rules and other exact relations like the Kramers-Kronig relations and the
Cauchy-Bunyakovsky-Schwarz inequalities. Unlike other theoretical approaches,
this method does not rely on a small parameter, is quite simple to use, and is
universal in the sense that it can be employed in non-Coulomb statistical systems as well \cite{He, dornheim6}. It is applicable to a wide class of plasma parameters in models of
one- and two-component plasmas, electron gas, etc. The basis of the method is,
of course, purely mathematical, and the specifics of the system are reflected
only in the moments; in this sense, the approach is model-free. Being
self-consistent, it expresses the dynamic properties of Coulomb
and non-Coulomb systems~\cite{Ortner_2000}, regardless of their geometry~\cite{arkhipov_cpp2014,PhysRevE.90.053102,PhysRevE.91.019903,arkhipov-etal.2017prl,arkhipov-etal.2020pre,Syzganbayeva_2022}, in terms
of their static characteristics, namely, the moments or the sum rules.

The coupling parameter used to describe classical systems is used in the
present work as a mere reference; it is the ratio of the potential interaction
energy of two elementary charges at a distance of the Wigner-Zeitz radius
$a=(3/4\pi n)^{1/3}$,  to the characteristic kinetic energy or the system temperature:%
\begin{equation}
\Gamma =(\beta e^{2})\, a^{-1}\ , \label{Gamma}%
\end{equation}
where $\beta =(k_{B}T)^{-1}$ is the inverse temperature in energy units;
$k_{B}$ and $n$ being the Boltzmann constant and the number density of the particles.

To characterize a quantum system we need two dimensionless parameters.
These are the degeneracy parameter $\theta$, %
\begin{equation}
\theta^{-1}=\beta E_{F}=1.84159\, \Gamma\cdot r_{s}^{-1}\ , \label{theta1}%
\end{equation}
which is the ratio of the system Fermi energy to its thermal energy, and the
density Brueckner parameter%
\begin{equation}
r_{s}=a/a_{B}=(ame^{2})/\hbar^{2}\ , \label{rs}%
\end{equation}
($a_{B}$ is the first Bohr radius). When these two parameters take values
around unity, we deal with the warm dense matter~\cite{dornheim_physrep_18}, more dilute quantum
systems are characterized by higher coupling with higher values of $r_{s}$~\cite{ara-etal.2021pop,Ara_2022,PhysRevB.107.195143}.

Consideration of the phenomenon of plasma stopping ability is usually
based on two theoretical approaches: i) the dielectric function formalism; ii) the
formalism of pair collisions. Here, the first approach is applied; cf. the plasma stopping is related to the longitudinal dielectric function of the medium~\cite{Lindhard_diss}:

\begin{equation}
S(\varv)\equiv -\frac{dE}{dx}=\frac{2(Ze)^{2}}{\pi \varv^{2}}\int_{0}^{\infty} \frac{dk}{k}\int_{0}%
^{k \varv}\omega\, [-\operatorname{Im} \epsilon^{-1}(k,\omega)]\, d\omega\ . \label{4}%
\end{equation}
Here $\epsilon(k,\omega)$ is the medium dielectric function, and $(Ze)$
and $\varv$\ are the charge and velocity of the incoming particle, $k$ is the
wavenumber. This expression contains a direct proportionality of the magnitude
of the polarizational energy loss to the square of the projectile charge,
thereby indicating the relationship between the linear
response and the first-order perturbation.

We can observe that the polarizational stopping power is determined by the (positive) loss
function, which we define here as%
\begin{equation}
L(k,\omega )=-\frac{\operatorname{Im}\epsilon^{-1}(k,\omega)}{\pi \omega}  \label{5}%
\end{equation}
so that
\begin{equation}
S(\varv)=\frac{2(Ze)^{2}}{\varv^{2}}\int_{0}^{\infty}\frac{dk}{k}\int_{0}^{k\varv}\omega^{2}\, L(k,\omega )\, d\omega\ . \label{6}%
\end{equation}

In this work, we take into account nine power moments of the loss function calculated as follows:%
\begin{equation}
\mu_s(k)=\int_{-\infty}^{\infty}\omega^sL(k,\omega)d\omega
\ ,\quad s=0,1,2,\ldots8. \label{7}%
\end{equation}

What matters is that these moments, if they converge, are strictly positive, at least  in
the physical context, and are known independently together with the
characteristic frequencies
\begin{equation}
\omega_{j}\left(  q\right)  =\sqrt{\frac{\mu_{2j}\left(  q\right)  }%
{\mu_{2j-2}\left(  q\right)  }}\ ,\quad j=1,2,3,4. \label{8}%
\end{equation}
Here and in what follows, we use the dimensionless wavenumber $q=ka$. The
calculations will be carried out with  the frequencies normalized to the plasma one,
 $\omega_{p}=\sqrt{4\pi ne^{2}/m}$.

\section{The mathematical background}\label{sec3}
The inverse dielectric function being a genuine response or Nevanlinna \cite{krein-book}
function is analytical in the (open) upper half-plane of the complex frequency $z=\omega+i\delta$. It possesses there a non-negative imaginary part, and satisfies the Kramers-Kronig relations,
\begin{equation}
\epsilon^{-1}(q,z)=1+\frac{1}{\pi}\int_{-\infty}^{\infty}\frac
{\operatorname{Im}\epsilon^{-1}(q,\omega )}{\omega-z}d\omega, \ \quad
z=\omega+i\delta\ ,\quad\delta>0.
\label{kk}%
\end{equation}
On the other hand, the Nevanlinna theorem~\cite{nevanlinna-book, krein-book, tkachenko-book} of the theory of moments
(Hamburger's moment problem)~\cite{shohat-book, akhiezer-book} establishes a one-to-one correspondence
between the Cauchy transform of the loss function and the unknown
phenomenological parameter function belonging to the same class of Nevanlinna
functions:%
\begin{equation}
\mathbf{\int}_{\mathbf{-\infty}}^{\mathbf{\infty}}\frac{L(q,\omega
)}{z-\omega }d\omega =\frac{E_{\ell+1}\left(  z;q\right)  +Q_{\ell}\left(
z;q\right)  E_{\ell}\left(  z;q\right)  }{D_{\ell+1}\left(  z;q\right)
+Q_{\ell}\left(  z;q\right)  D_{\ell}\left(  z;q\right)  }\ . \label{nt}%
\end{equation}
In addition, the function $Q_{\ell}\left(  z;q\right)  $, which we call the Nevanlinna parameter
function (NPF), must be such that as $z \rightarrow \infty$ in any angular sector $\vartheta < arg\, z< \pi - \vartheta$  ($ 0 < \vartheta < \pi$)
in the upper half-plane $\operatorname{Im}z=\delta > 0$, it decays faster than $1/z$. It is obvious that NPF cannot be determined within the method of moments, and we need
a physical considerations to model it. The coefficients of the linear-fractional
transformation (\ref{nt}) are orthogonal, with the weight $L(q,\omega)$, polynomials $D_{\ell}\left(
z;q\right)  $ and their conjugate ones are%
\[
E_{\ell}\left(  z;q\right)  =\mathbf{\int}_{\mathbf{-\infty}}^{\mathbf{\infty
}}\frac{D_{\ell}(q,z)-D_{\ell}(q,\omega )}{z-\omega } L(q,\omega ) d\omega \ ,\;
\ell=0,1,\ldots 5.
\]

It is easy to establish a bridge between the inverse dielectric function and the NPF:%
\begin{align}
\epsilon^{-1}(q,z)  &  =1-\int_{-\infty}^{\infty}\frac{\omega\, L(q,\omega
)}{\omega-z}d\omega =\label{r1}\\
&  =1-\mu_0\left(  q\right)  +z\frac{E_{\ell+1}\left(  z;q\right)  +Q_{\ell
}\left(  z;q\right)  E_{\ell}\left(  z;q\right)  }{D_{\ell+1}\left(
z;q\right)  +Q_{\ell}\left(  z;q\right)  D_{\ell}\left(  z;q\right)  }\ .
\label{r}%
\end{align}
The standard Gram-Schmidt procedure allows one to express the above polynomials
in terms of frequencies (\ref{8}), see Sec.~7.%

We have taken into account that the odd-order moments vanish due to the parity of
the loss function. It immediately follows from the Kramers-Kronig relations (\ref{kk}) or (\ref{r1})-(\ref{r}) that the lowest order moment $\mu_0\left(  q\right)  $ is determined by the zero-frequency value of the
(inverse) dielectric function,
\[
\epsilon^{-1}(q,0)=\lim_{\delta\downarrow0}\epsilon^{-1}(q,i\delta)\ :
\]%
\begin{equation}
\mu_0\left(  q\right)  =1-\epsilon^{-1}(q,0)\ . \label{9}%
\end{equation}
The rest of the involved moments determine the asymptotic expansion of
(\ref{r1}) along any ray in the upper half-plane $\operatorname{Im}z>0$\ from
the origin to infinity,
\[
\epsilon^{-1}(q,z\rightarrow\infty)\simeq1+\mu_0\left(  q\right)  \left(
\frac{\omega_{1}^{2}}{z^{2}}+\frac{\omega_{1}^{2}\omega_{2}^{2}}{z^{4}}%
+\frac{\omega_{1}^{2}\omega_{2}^{2}\omega_{3}^{2}}{z^{6}}+\frac{\omega_{1}%
^{2}\omega_{2}^{2}\omega_{3}^{2}\omega_{4}^{2}}{z^{8}}\right)
\]
\begin{equation}
 +O\left(
z^{-9}\right)  \ . \label{r1x}%
\end{equation}
The same expansion can be obtained from (\ref{r}):%
\[
\epsilon^{-1}(q,z\rightarrow\infty)\simeq1+\mu_0\left(  q\right)  \left(
\frac{\omega_{1}^{2}}{z^{2}}+\frac{\omega_{1}^{2}\omega_{2}^{2}}{z^{4}}%
+\frac{\omega_{1}^{2}\omega_{2}^{2}\omega_{3}^{2}}{z^{6}}+\frac{\omega_{1}%
^{2}\omega_{2}^{2}\omega_{3}^{2}\omega_{4}^{2}}{z^{8}}+\right.
\]%
\begin{equation}
\left.  +C_{0}\left(  q\right)  \mathcal{A}\left(  q\right)  \frac
{Q_{4}\left(  z;q\right)  }{z^{9}}\right)  +\mathcal{O}\left(  z^{-10}\right)
\ ,
\end{equation}
where
\[
\mathcal{A}\left(  q\right)  =\omega_{1}^{2}\omega_{2}^{2}\left(  \omega
_{3}^{2}\omega_{4}^{2}+\frac{\omega_{1}^{2}\omega_{2}^{2}\left(  \omega
_{3}^{2}-\omega_{2}^{2}\right)  -\omega_{2}^{2}\omega_{3}^{2}\left(
\omega_{3}^{2}-\omega_{1}^{2}\right)  }{\omega_{2}^{2}-\omega_{1}^{2}}\right)
\ .
\]
The intrinsically complex correction of the order $z^{-9}$ implies that the
Nevanlinna linear-fractional representation of the inverse dielectric
function~(\ref{r}) guarantees the verification of the moment sum rules
$\left\{  \mu_{2},0,\mu_{4}\left(  q\right)  ,0,\mu_{6}\left(  q\right)
,0,\mu_{8}\left(  q\right)  \right\}  $ automatically, independently of our
mathematically correct model of the NPF $Q_{4}\left(  z;q\right)  $, which is
set here to be frequency-independent:%
\[
Q_{4}\left(  z;q\right)  =Q_{4}\left(  0;q\right)  =ih_{4}\left(  q\right)
\ .
\]

\noindent The moments and additional details of our NPF model are provided in the Appendix.

Notice that the $f-$sum rule%
\begin{equation}
\mu_2=\omega_{p}^{2}=\frac{3e^{2}}{ma^{3}} \label{c2}%
\end{equation}
reduces the expansion (\ref{r1x}) to the familiar form,%
\begin{equation}
\epsilon^{-1}(q,z\rightarrow\infty)\simeq1+\frac{\omega_{p}^{2}}{z^{2}}\left(
1+\frac{\omega_{2}^{2}}{z^{2}}+\frac{\omega_{2}^{2}\omega_{3}^{2}}{z^{4}%
}+\frac{\omega_{2}^{2}\omega_{3}^{2}\omega_{4}^{2}}{z^{6}}\right)  +O\left(
z^{-9}\right)  \ . \label{ff}%
\end{equation}

Finally, to point out that the Cauchy-Bunyakovsky-Schwarz inequality
explicitly implies that for any wavenumber the characteristic frequencies satisfy the inequalities
\begin{equation}
\omega_{1}\left(  q\right)  \leq\omega_{2}\left(  q\right)  \leq\omega
_{3}\left(  q\right)  \leq\omega_{4}\left(  q\right)  \ ,\label{cbs}%
\end{equation}
and that the equality options here are obviously never satisfied in physical
systems. Nevertheless, these inequalities guarantee the possibility of the
realization of the exact simple asymptotic transformation of the nine-moment
construction in~(\ref{r}) into the five-moment one by the following successive limiting steps:
$\omega_{4}\left(  q\right)/\omega_{3}\left(  q\right)%
\rightarrow\infty$,
followed by
$\omega_{3}\left(  q\right)/\omega_{2}\left(  q\right)
\rightarrow\infty.
$

\section{Numerical results}\label{sec4}

\begin{figure}[]
\includegraphics[width=0.5\textwidth]{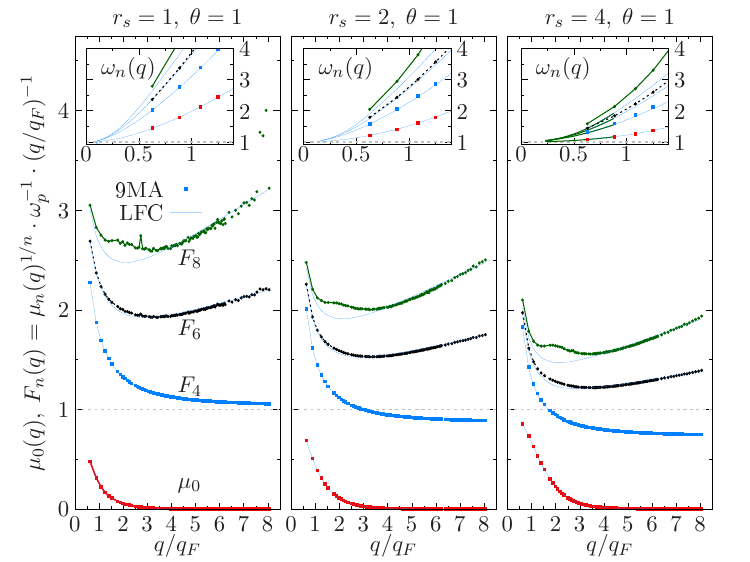}
\includegraphics[width=0.5\textwidth]{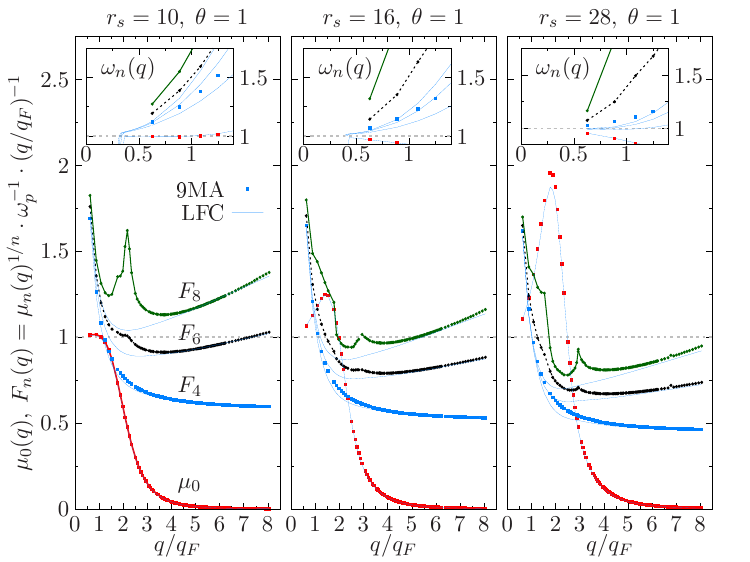}

\vspace{-0.cm}
\caption{The wavenumber dependence of the rescaled power moments $F_n(q)$ and $\mu_0(q)$ estimated in the 9-moment approach: $\{\mu_0, F_4\}$ by QMC and $\{F_6, F_8\}$ by Shannon entropy maximization. The insert figures demonstrate the behavior in the long wavelength limit. For comparison, the ESA results~\cite{ESA, PhysRevB.103.165102} combined with the QMC-LFC ($r_s\geq 22$) (solid blue lines) are provided (cited in the text as LFC).
For $r_s\gtrsim 16 $ formation of the roton mode \cite{Kalman_2010}, which can be identified as a local minimum in $F_8(q)$ and $F_6(q)$ within the wavenumber interval $1.5 \leq  q/q_F \leq 3$ (cf. the panels with $r_s=16; 28$), was explicitly revealed in ~\cite{PhysRevB.107.195143}.}
\label{fig:PowerM}
\end{figure}


We have carried out extensive calculations of the polarizational stopping
power of a uniform moderately and strongly coupled electronic gas as a function
of the projectile velocity $\varv$ and the dimensionless parameters $\{r_{s},
\theta\}$.
The expression for the loss function is obtained directly from~(\ref{r}) and the NPF
model as described in Sec.~7:%
\begin{equation}
L\left(  \omega;q\right)  =\frac{\omega_{p}^{2}R_{2}\left(  \omega_{2}%
^{2}-\omega_{1}^{2}\right)  }{\left(  \omega\left(  \omega^{2}-\omega_{2}%
^{2}\right)  +R_{1}\left(  \omega^{2}-\omega_{1}^{2}\right)  \right)
^{2}+R_{2}^{2}\left(  \omega^{2}-\omega_{1}^{2}\right)  ^{2}} .\label{L}%
\end{equation}
We note that this loss function satisfies the nine sum rules automatically, by construction, according to the Nevanlinna theorem~\cite{krein-book}.

The fundamental ingredient of this approach, cf.
the characteristic frequencies $\omega_{1(2)}(q)$ [defined by the power moments $\mu_0(q),\mu_2,\mu_4(q)$] have been provided by the fermionic QMC code~\cite{PhysRevB.107.195143}, while the higher-order  frequencies $\omega_{3(4)}(q)$ were determined by the Shannon information entropy maximization procedure; see Sec.~\ref{w3w4} for details.

Our numerical results (filled symbols) are displayed in Fig.~1 along with the ESA data~\cite{ESA, PhysRevB.103.165102} for $r_s \leq 20$ (solid blue lines). The latter have been obtained in our calculations using the machine learning representation of the static local field correction (LFC) for UEG~\cite{dorn.2019jcp}.
For larger $r_s (\geq 22)$ we use the static local field correction reconstructed from the QMC data~\cite{PhysRevB.107.195143}, $G(q)=1-q^2/4\pi [\chi_0^{-1}(q)-\chi^{-1}(q)]$, and the corrected short-wavelength limiting form, $\lim\limits_{q\rightarrow \infty} G(q)=1-g(0)$, with $g(0)$ being the on-top pair distribution function~\cite{Tanaka1986}.

For better visibility, the moments $\mu_n(q)$ ($n\geq 4$) are rescaled to $F_n(q)=\mu_n^{1/n}(q)\, \omega_p^{-1} \, (q/q_F)^{-1}$. This allows us to accurately analyze the discrepancies between our approach and the LFC theory. The latter, in general, does not guarantee the correct values of spectral moments with $n\geq 4$ since the electron correlations are treated only in the static approximation equivalent to the introduction of an effective interaction potential.

As the value of the density parameter $r_s$ increases (from left to right in Fig.~\ref{fig:PowerM}) the values of the higher moments approach each other. The Cauchy-Bunyakovsky-Schwarz inequalities~(\ref{cbs}) remain always satisfied, even at the lowest available wave numbers (see inserts). The 9MA power moments are labeled successively: zero (red), fourth (blue), sixth (black), and eighth (green). The LFC predictions are indicated by the solid blue lines.

First, we can validate that both theories demonstrate remarkable agreement at moderate to high densities ($4\geq r_s\geq 1$) for the lowest moments $\mu_n(q)$ ($n\leq 4$). However, as we approach the dilute phase ($r_s\geq 10$), some noticeable deviations are always present in $F_8(q)$ and systematically arise in $F_6(q)$, becoming less pronounced in $F_4(q)$. The physical reason for this discrepancy has been analyzed in detail in~\cite{PhysRevB.107.195143} and should be attributed to a bimodal structure of the excitation spectrum consisting of a low-frequency roton-like and a high-frequency branches exactly in the same interval $1.5 \leq  q/q_F \leq 3$, where the deviations in $F_n(q)$ between both theories are the largest. Finally, we note that by its construction, the LFC should accurately satisfy the zero-order moment $\mu_0(q)$, and from our data we confirm this property.


\begin{figure}[]
\includegraphics[width=0.5\textwidth]{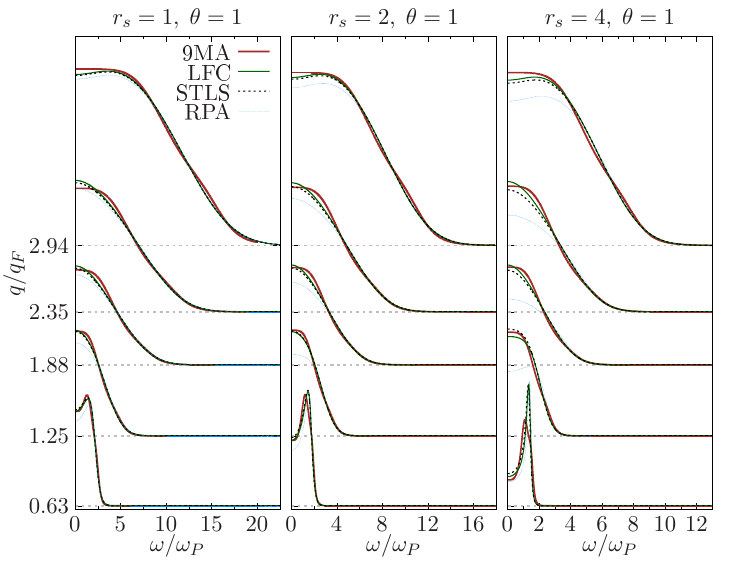}
\includegraphics[width=0.5\textwidth]{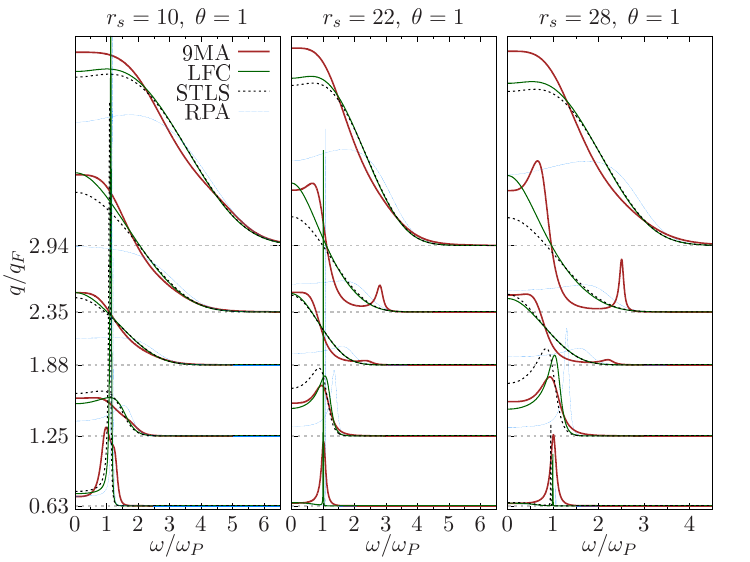}
\vspace{-0.3cm}
\caption{The frequency dependence of the loss function $L(\omega; q)$, Eq.~(\ref{L}),
estimated in different approximations (RPA, STLS, LFC: ESA ($r_s\leq 20$) $\&$ QMC-LFC ($r_s\geq 22$) and the 9-moment approach).
The dashed horizontal lines specify the selected wavenumbers in the
range $0.63 \leq q/q_F\leq 2.94$. With the increase of the coupling strength
(larger $r_s$ values) all theories predict the formation of a sharp plasmon resonance
for $q/q_F\leq 0.63$.}
\label{fig:LossF0rs1-rs4}
\end{figure}

Next, in Fig.~\ref{fig:LossF0rs1-rs4} we demonstrate how different theories resolve the frequency dependence of the loss function $L(q,\omega)$. This spectral density is a central quantity which defines both the stopping power and the dynamic structure factor.

The RPA theory, which neglects the electronic correlation effects, shows its inconsistency for $L(q,\omega)$ already at moderate coupling ($r_s\sim 4$) and its results deviate significantly from those of more sophisticated theoretical approaches. The STLS~\cite{STLS} and LFC include correlation effects in the static approximation and demonstrate quite a reasonable mutual agreement for $r_s \leq 10$.

Now we compare these results to our 9MA reconstruction which satisfies a wide class of 
exact relations and physical limitations known either analytically or from \textit{ab initio} QMC simulations~\cite{PhysRevB.107.195143}.

At moderate couplings ($r_s\leq 10$) the observed discrepancies with STLS and LFC are only quantitative; they do not lead to qualitative differences in the form of the loss function. However, the higher coupling ($r_s \geq 16$, $\theta=1$) significantly affects the spectrum: we predict that the projectiles interact with two modes or quasiparticles revealed in the bimodal spectrum of density fluctuations~\cite{PhysRevB.107.195143} in distinct ways, and two maxima appear
instead of the traditional unique one located near the thermal velocity of electrons.

\begin{figure}[]
\vspace{-0.2cm}
\centering\includegraphics[width=0.5\textwidth]{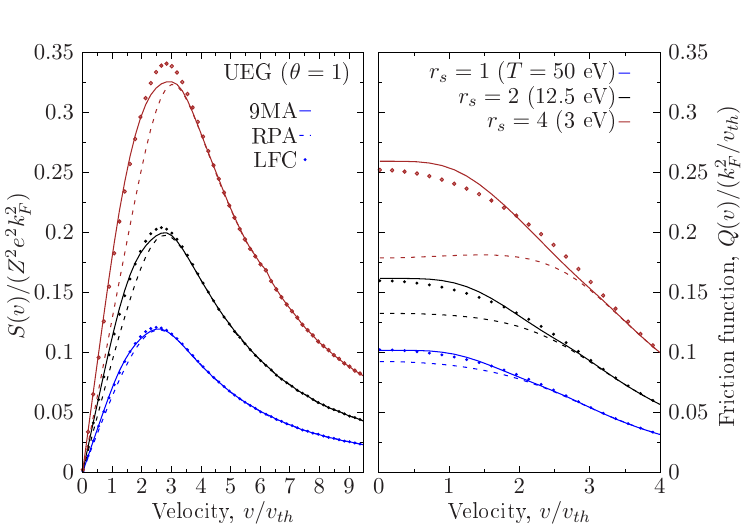}
\vspace{-0.3cm}
\caption{{\it Left:} Stopping power at $\theta=1$ and $r_s=1,2$ and $4$ vs. the ion velocity. {\it Right:} Friction function $Q(\varv)$. At small velocities,
$\varv/\varv_{th} < 1$, the stopping power is linear in the ion velocity. The label indicates the absolute temperature of UEG in eV estimated for a given $\{r_s,\theta=1\}$-combination.}
\label{fig:S124}
\end{figure}

\begin{figure}[]
\vspace{-0.2cm}
\centering\includegraphics[width=0.5\textwidth]{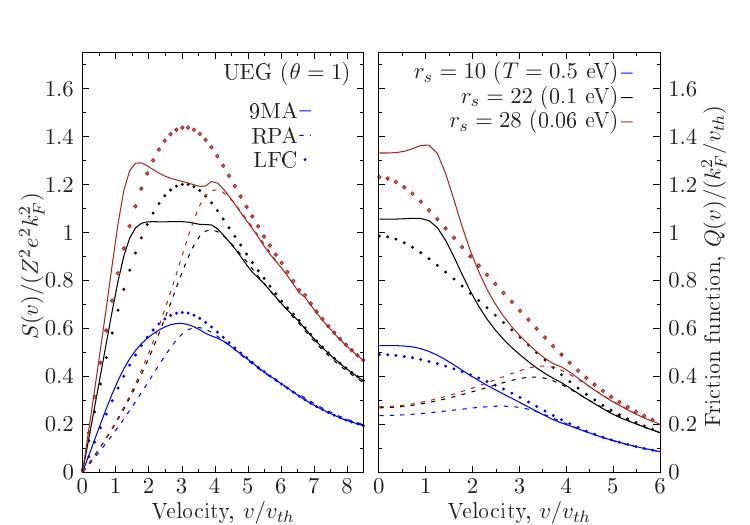}
\vspace{-0.3cm}
\caption{{\it Left:} Stopping power at $\theta=1$ and $r_s=10,22$ and $28$ vs. the ion velocity. {\it Right:} Friction function $Q(\varv)$. At small velocities,
$\varv/\varv_{th} < 1$, the stopping power remains linear in the ion velocity (which is physically reasonable) only within the 9MA and the RPA theories.}
\label{fig:S28}
\end{figure}

The results for the loss function in Fig.~\ref{fig:LossF0rs1-rs4} have been used to evaluate the stopping power $S(\varv)$ via Eq.~(\ref{6}). The latter is presented in Fig.~\ref{fig:S124} and Fig.~\ref{fig:S28}.
As before, we conducted three types of calculations for 9MA, RPA, and the LFC-based approach~\cite{dorn.2019jcp}.

In addition to the stopping power we have estimated the friction function (see the right panels in Fig.~3,~4)
\begin{equation}
Q(\varv)=\frac{S(\varv)}{(Ze)^{2} \, \varv }. \label{fr}%
\end{equation}


First, in the WDM regime ($\theta=1$ and $r_s=1,2,4$)
the derived conclusions are similar to those provided for the loss function data. In fact, we performed calculations of both the stopping power and the friction function using the 9MA and LFC approaches and observed good quantitative agreement with the results of ~\cite{mold.2020pre}. The asymptotic behavior at low and high projectile velocity nearly coincides, and some minor deviations are revealed only near the maximum of the stopping curve, but they increase systematically in the velocity interval $2 \leq \varv/\varv_{th}  \leq  3.5$.
In general, we find that the LFC overestimates the energy losses around the maximum, and underestimates the friction coefficient for projectile velocities below the thermal gas velocity ($\varv\lesssim \varv_{th}$).

Next, we analyze the dependence of the stopping power velocity at higher couplings ($\theta=1$ and $r_s=10,22,28$), when there is a significant discrepancy between the predictions of 9MA and LFC due to different spectra of density fluctuations~\cite{PhysRevB.107.195143}.
Precisely, though the stopping asymptotic behavior is maintained at low and high projectile velocities, the two-peak structure of the absorption spectrum (present in two right most panels in Fig.~\ref{fig:LossF0rs1-rs4}) substitutes the well-known stopping peak at roughly $3 \varv_{th}$: this unique Bragg peak first converts into a shelf between approximately $2 \varv_{th}$ and $4 \varv_{th}$, and then it splits into two lower peaks. This fine structure of the stopping power is reflected also in a novel behavior of the friction function: i) at $r_s=22 (28)$ it maintains the limiting low-velocity value up to the thermal velocity; ii) at $r_s=28$ we even observe how the friction function, instead of the traditional decrease, is slightly enhanced when the projectile velocity approaches the thermal velocity of the electrons. This novel effect is not captured by the LFC, and, moreover, the LFC scheme in this regime does not conserve the linear scaling between the energy losses and the projectile velocity. In contrast, the RPA conserves this linear scaling, but strongly underestimates the stopping power for $\varv \leq 4.5 \varv_{th}$. Interestingly enough, at higher velocities, the 9MA and the RPA demonstrate a nearly perfect agreement for $S(\varv)$ and $Q(\varv)$, and both deviate from the LFC. This important observation can be better resolved in Fig.~\ref{fig:S28}.

\vspace{0.cm}




\section{Conclusions}\label{sec5}
The present novel results are based on two recent developments: the fermionic propagator path integral quantum Monte-Carlo method and the nine-moment version of the self-consistent method of moments with both accomplishments described in detail in~\cite{PhysRevB.107.195143}, see also the corresponding references cited therein. These improvements have allowed us to reveal the dynamic properties of the UEG under strong coupling conditions, i.e., at the Bruckner parameter values $r_s=a/a_B \gtrsim 10$. New features of the low density electron gas stopping power together with the friction function directly reflect our recently disclosed bimodal structure of the density fluctuation spectrum for $r_s\geq 16$. The latter manifests itself as a shelf-like structure near the stopping power maximum, and a nearly constant friction coefficient $Q(\varv)$ for $\varv \leq \varv_{th}$.

We suggest that this new structure of the UEG stopping power and friction function projectile velocity dependence must be accompanied by other peculiarities of the UEG dynamic properties. But the investigation of these novel features goes beyond the scope of the present work.

Majority of currently available alternative computational approaches to the analysis of the UEG dynamic properties intrinsically violate all the sum rules accounted for here, except the $f-$sum rule. 
Electrons in real strongly coupled Coulomb liquids play decisive role in their fast-projectile stopping capacity \cite{Hentschel_2023}. Thus, we hope that future direct measurements in dense plasmas would qualitatively confirm our predictions of both theoretical and practical importance.

Preliminary results in this direction might be obtained within the DFT approach, see the recent developments presented in~\cite{ward_2024,white_98,ziegler_10,schleife_15,kononov_2023,nazarov_2005,Kononov_24,sarasola_04} and ~\cite{echenique_1991}.

\section*{Acknowledgments}
The authors acknowledge the support via the grants AP09260349 of the
Ministry of Education and Science, Kazakhstan and PID2021-127381NB-I00 of the Ministry of Science and Innovation, Spain.

\section*{Appendix: Spectral moments}
\subsection{The moment $\mu_{4}\left(  q\right)  $.}

The explicit expression for this sum rule has been known since long~\cite{puff}. It can be
obtained, e.g., directly from the system Hamiltonian within the theory of linear
response and using the second quantization procedures~\cite{tkachenko-book}. The fourth moment,%
\begin{eqnarray}
&&\mu_4\left(  q\right)  =\omega_{p}^{4}\left\{  1+\frac{2q^{2}}{3\Gamma}%
\beta\left\langle E_{\mathrm{kin}}\right\rangle +\frac{q^{4}}{12r_{s}%
}+U\left(  q\right)  \right\}  \ ,\label{c4}\\
&&U\left(  q\right)  =\frac{1}{3\pi}\int_{0}^{\infty}\left[  S\left(  p\right)
-1\right]  f\left(  p,q\right)  p^{2}dp\ ,\label{c41}%
\end{eqnarray}
contains the average interacting-system kinetic energy $\left\langle
E_{\mathrm{kin}}\right\rangle $ reduced in the case of an ideal gas to the
Fermi integral $I_{3/2}(\eta)$: $\left\langle E_{\mathrm{kin}}\right\rangle
=3\theta^{3/2}I_{3/2}\left(  \eta\right)  /2\beta$, the one-particle
contribution $q^{4}/12r_{s}$, and the correlation contribution $U\left(
q\right)  $ determined by the system static structure factor $S(q)$ provided,
e.g., by the \textit{ab initio} QMC simulations~\cite{dornheim.2018prl,PhysRevB.107.195143}. The angular averaging in the
momentum vector is reflected by the factor%
\begin{equation}
f\left(  p,q\right)  =\frac{5}{6}-\frac{p^{2}}{2q^{2}}+\frac{\left(
p^{2}-q^{2}\right)  ^{2}}{4q^{3}p}\ln\left\vert \frac{q+p}{q-p}\right\vert
\ .\label{44}%
\end{equation}
It follows from our expression for the loss function that if we neglect the
energy dissipation, the dispersion of the plasma optical collective mode will
be determined by the fourth moment (\ref{c4}) which, in the long-wavelength
limiting case reproduces the ideal gas Vlasov dispersion form and in the
warm-dense regime provides the rich picture of the plasma collective modes
deduced in \cite{PhysRevB.107.195143}. 

\subsection{The frequencies $\omega_{3}^{2}\left(  q\right)  $ and $\omega_{4}^{2}\left(  q\right)$.}\label{w3w4}

We understand that the higher-order non-vanishing moments and the frequencies
$\omega_{3}(q)$ and $\omega_{4}(q)$ are determined by the virtually unknown
three- and four-body static correlation functions. As \textit{ab initio} QMC data for
them are not yet available, we determine them here by means of the Shannon
information entropy maximization procedure~\cite{shannon.1948,khinchin.53umn,zubarev-book}. Precisely, we employ the
two-parameter Shannon entropy functional:%
\begin{equation}
\mathfrak{\Sigma}\left(  \omega_{3},\omega_{4};q\right)  =-%
{\displaystyle\int\limits_{-\infty}^{\infty}}
L\left(  q,\omega\right)  \ln\left[  L\left(  q,\omega\right)  \right]
d\omega\ .\label{sh0}%
\end{equation}
Since the loss function in the nine-moment approximation depends on the
additional characteristic frequencies $\omega_{3}(q)$ and $\omega_{4}(q)$, we
find the entropy critical points from the following system of equations,%
\begin{equation}
\left\{
\begin{array}
[c]{c}%
-%
{\displaystyle\int\limits_{-\infty}^{\infty}}
\left\{  \frac{\partial L\left(  q,\omega\right)  }{\partial\omega_{3}}%
\ln\left[  eL\left(  q,\omega\right)  \right]  \right\}  d\omega=0\ ,\\
-%
{\displaystyle\int\limits_{-\infty}^{\infty}}
\left\{  \frac{\partial L\left(  q,\omega\right)  }{\partial\omega_{4}}%
\ln\left[  eL\left(  q,\omega\right)  \right]  \right\}  d\omega=0\ ,
\end{array}
\right.  \label{sh1}%
\end{equation}
using the Newton-Raphson method. In the gradient descent, as the starting
points $\{\omega_{30}(q),\omega_{40}(q)\}$ we chose the ratios of the
corresponding frequency moments of an ideal Fermi gas distribution:%
\[
\omega_{30}^{2}\left(  q\right)  =\frac{I_{5/2}(\eta)}{I_{3/2}(\eta)}%
\omega_{20}^{2}\left(  q;\eta\right)  \ ,\quad\omega_{40}^{2}\left(  q\right)
=\frac{I_{7/2}(\eta)}{I_{5/2}(\eta)}\omega_{20}^{2}\left(  q;\eta\right)  \ ,
\]
where%
\begin{align*}
\omega_{20}^{2}\left(  q;\eta\right)   &  =\frac{I_{1/2}(\eta)}{I_{3/2}(\eta
)}\omega_{2}^{2}\left(  q\right)  \ ,\quad I_{\ell}(\eta)=\int_{0}^{\infty
}\frac{x^{\ell}dx}{1+\exp\left(  x-\eta\right)  },\\
\ell &  =\frac{1}{2},\frac{3}{2},\frac{5}{2},\frac{7}{2}.
\end{align*}
The Hessian of the entropy (\ref{sh0}) was studied to warrant its maximization
at the solution of the system (\ref{sh1}). 

\subsection{The Nevanlinna parameter
function model.}\label{m}

It was shown~\cite{arkhipov-etal.2017prl, arkhipov-etal.2020pre} that in classical Coulomb and Yukawa dense
plasmas within the five-moment version of the self-consistent moment approach
the static approximation for the Nevanlinna function,
\begin{equation}
Q_{2}\left(  z;q\right)  =Q_{2}\left(  z;q\right)  =ih_{2}\left(  q\right)
\ ,\label{sa}%
\end{equation}
with $h_{2}\left(  q;\omega_{1},\omega_{2}\right)  =\omega_{2}^{2}\left(
q\right)  /\left(  \sqrt{2}\omega_{1}\left(  q\right)  \right)  $ was
sufficient to describe dynamic properties of those statistical systems. The latter value for the
static Nevanlinna parameter $h_{2}\left(  q;\omega_{1},\omega_{2}\right)  >0$
was obtained~\cite{arkhipov-etal.2017prl} from the observation that the zero-frequency
Rayleigh mode was absent in those systems. It is curious enough that in the
electron gas we deal with here, this mode is equally absent. Nevertheless, an accurate description of complicated collective processes in a UEG is
possible only beyond the static local-field approximation~\cite{ESA, PhysRevB.103.165102}, as it was demonstrated in detail in~\cite{PhysRevB.107.195143}. To transfer this methodology to the stopping power calculation, we adopt here the nine-moment Nevanlinna formula as well, and
equalize the five-moment and nine-moment expressions for the dielectric
function:%
\[
\frac{E_{3}\left(  z;q\right)  +Q_{2}\left(  z;q\right)  E_{2}\left(
z;q\right)  }{D_{3}\left(  z;q\right)  +Q_{2}\left(  z;q\right)  D_{2}\left(
z;q\right)  }=\frac{E_{5}\left(  z;q\right)  +Q_{4}\left(  z;q\right)
E_{4}\left(  z;q\right)  }{D_{5}\left(  z;q\right)  +Q_{4}\left(  z;q\right)
D_{4}\left(  z;q\right)  }.
\]
This leads to a direct linear-fractional expression for the five-moment NPF in
terms of the nine-moment one. The coefficients of this transformation are the orthogonal polynomials provided below. Then, by virtue of the same
physically motivated consideration~\cite{arkhipov-etal.2017prl}, we employ the static approximation for
the 
nine-moment NPF: 
\begin{equation}
Q_{4}\left(  z;q\right)=Q_{4}\left(  0;q\right)=h_{4}\left(  q\right).%
\end{equation}
Thus we constructed the dynamic five-moment Nevanlinna function:
\begin{eqnarray}%
&&Q_{2}\left(  q,\omega\right)  =-\frac{\Omega_{3}^{2}\left(  \omega
+ih_{4}\right)  }{\omega\left(  \omega+ih_{4}\right)  -\Omega_{4}^{2}}%
=R_{1}+iR_{2}\ ,\\
&&h_{4}\left(  q\right)  =\frac{\omega_{3}^{2}\left(  \omega_{4}^{2}-\omega
_{3}^{2}\right)  \left(  \omega_{2}^{2}-\omega_{1}^{2}\right)  }{\omega
_{1}\sqrt{2\left(  \omega_{3}^{2}-\omega_{1}^{2}\right)  \left(  \omega
_{3}^{2}-\omega_{2}^{2}\right)  ^{3}}}\ .
\label{Q24}%
\end{eqnarray}
Notice that the Cauchy-Bunyakovsky-Schwarz inequalities (\ref{cbs}) guarantee
the fulfillment of the required properties of the Nevanlinna and the inverse
dielectric functions. 


For completeness, we provide here the explicit expressions for the orthogonal polynomials as well:%

\[
D_{2}\left(  z;q\right)  =z^{2}-\omega_{1}^{2}\ ,\quad D_{3}\left(
z;q\right)  =z\left(  z^{2}-\omega_{2}^{2}\right)  \ ,
\]%
\begin{align*}
D_{4}\left(  z;q\right)   &  =z^{4}+a_{2}\left(  q\right)  z^{2}+a_{4}\left(
q\right)  \ ,\\
D_{5}\left(  z;q\right)   &  =z\left(  z^{4}+c_{1}\left(  q\right)
z^{2}+c_{3}\left(  q\right)  \right)  \ ;
\end{align*}%
\begin{align*}
E_{2}\left(  z;q\right)   &  =\mu_0\left(  q\right)  z\ ,\\
E_{3}\left(  z;q\right)   &  =\mu_0\left(  q\right)  \left[  z^{2}-\left(
\omega_{2}^{2}-\omega_{1}^{2}\right)  \right]  \ ,
\end{align*}%
\begin{align*}
E_{4}\left(  z;q\right)   &  =\mu_0\left(  q\right)  \left(  z^{3}%
+b_{3}\left(  q\right)  z\right)  \ ,\\
E_{5}\left(  z;q\right)   &  =\mu_0\left(  q\right)  \left(  z^{4}%
+d_{2}\left(  q\right)  z^{2}+d_{0}\left(  q\right)  \right)  \ ;
\end{align*}
where%
\begin{align*}
a_{2}  &  =-\omega_{2}^{2}\frac{\omega_{1}^{2}-\omega_{3}^{2}}{\omega_{1}%
^{2}-\omega_{2}^{2}}\ ,\quad a_{4}=\omega_{1}^{2}\omega_{2}^{2}\frac
{\omega_{2}^{2}-\omega_{3}^{2}}{\omega_{1}^{2}-\omega_{2}^{2}}\ ,\\
b_{3}  &  =-\frac{\omega_{3}^{2}\omega_{2}^{2}-2\omega_{2}^{2}\omega_{1}%
^{2}+\omega_{1}^{4}}{\omega_{2}^{2}-\omega_{1}^{2}}\ ;\\
c_{1}  &  =-\frac{\omega_{2}^{2}\omega_{3}^{2}-\omega_{3}^{2}\omega_{4}^{2}%
}{\omega_{2}^{2}-\omega_{3}^{2}}\ ,\quad c_{3}=\omega_{2}^{2}\omega_{3}%
^{2}\frac{\omega_{3}^{2}-\omega_{4}^{2}}{\omega_{2}^{2}-\omega_{3}^{2}}\ ;\\
d_{2}  &  =\frac{\omega_{1}^{2}\left(  \omega_{2}^{2}-\omega_{3}^{2}\right)
+\omega_{3}^{2}\left(  \omega_{4}^{2}-\omega_{2}^{2}\right)  }{\omega_{2}%
^{2}-\omega_{3}^{2}}\ ,\\
d_{0}  &  =\omega_{1}^{2}\omega_{2}^{2}+\omega_{3}^{2}\frac{\omega_{1}%
^{2}\left(  \omega_{4}^{2}-\omega_{2}^{2}\right)  +\omega_{2}^{2}\left(
\omega_{3}^{2}-\omega_{4}^{2}\right)  }{\omega_{2}^{2}-\omega_{3}^{2}}\ ,
\end{align*}
and for other parameters of the NPF we employ:

\begin{align}
R_{1}  &  =\frac{-\omega\Omega_{3}^{2}\left(  \omega^{2}-\Omega_{4}^{2}%
+h_{4}^{2}\right)  }{\left(  \omega^{2}-\Omega_{4}^{2}\right)  ^{2}+\left(
\omega h_{4}\right)  ^{2}}\ ,\\
R_{2}  &  =\frac{\Omega_{3}^{2}\Omega_{4}^{2}h_{4}}{\left(  \omega^{2}%
-\Omega_{4}^{2}\right)  ^{2}+\left(  h_{4}\omega\right)  ^{2}}\ ,\\
\Omega_{3}^{2}  &  =\frac{\omega_{2}^{2}\left(  \omega_{3}^{2}-\omega_{2}%
^{2}\right)  }{\omega_{2}^{2}-\omega_{1}^{2}}>0,\\
\Omega_{4}^{2}  &  =\frac{\omega_{3}^{2}\left(  \omega_{4}^{2}-\omega_{3}%
^{2}\right)  }{\left(  \omega_{3}^{2}-\omega_{2}^{2}\right)  }-\frac
{\omega_{1}^{2}\left(  \omega_{3}^{2}-\omega_{2}^{2}\right)  }{\left(
\omega_{2}^{2}-\omega_{1}^{2}\right)  }>0.
\end{align}

\bibliography{ref}

\end{document}